\documentclass[%
 reprint,showpacs,showkeys,nofootinbib,amsmath,amssymb, aps,]{revtex4-1}

\usepackage{graphicx}
\usepackage{dcolumn}
\usepackage{bm}

\usepackage[dvips]{color}

\begin{document}

\title{Big Bang nucleosynthesis in visible and hidden-mirror sectors.}

\author{Paolo Ciarcelluti}
 \email{paolo.ciarcelluti@gmail.com}
 \affiliation{Web Institute of Physics, www.wiph.org}

\date{\today}

\begin{abstract}

One of the still viable candidates for the dark matter is the so-called mirror matter.
Its cosmological and astrophysical implications were widely studied in many aspects, pointing out the importance to go further with research and refine the studies.
In particular, the Big Bang nucleosynthesis provides a strong test for every dark matter candidate, since it is well studied and involves relatively few free parameters. 
The necessity of accurate studies of primordial nucleosynthesis with mirror matter has then emerged.
In order to fill this lack, I present here the results of accurate numerical simulations of the primordial production of both ordinary nuclides and nuclides made of mirror baryons, in presence of a hidden mirror sector with unbroken parity symmetry and with gravitational interactions only.
These elements are the building blocks of all the structures forming in the Universe, therefore their chemical composition is a key ingredient for astrophysics with mirror dark matter.
The production of ordinary nuclides show differences from the standard model for a ratio of the temperatures between mirror and ordinary sectors $x = T'/T \gtrsim 0.3$, and they present an interesting decrease of the abundance of ${\rm ^7Li}$.
For the mirror nuclides, instead, one observes an enhanced production of ${\rm ^4He}$, that becomes the dominant element for $x  \lesssim 0.5$, and much larger abundances of heavier elements.

\end{abstract}



\maketitle


\section{Introduction}

The nature of the dark matter of the Universe is still completely unknown, and mirror matter represents one of the possible promising candidates.
Its postulation derives theoretically from the necessity to restore the parity symmetry of the physical laws, and phenomenologically from the need of describing the physics at any scales, from the whole cosmos to the elementary particles.
In its basic theory, mirror matter is formed by baryons with exactly the same properties as our ordinary baryons, but with opposite handedness (right) of weak interactions, so that globally the system of all particles together (ordinary and mirror) is parity symmetric.
All the particles and the coupling constants are the same, then the physical laws of mirror matter are the same as that of ordinary matter, but the only interaction between the two kinds of particles is gravitational, while the other fundamental interactions act separately in each sector.
This is valid also for electromagnetic interactions, meaning that a mirror photon would interact with mirror baryons but not with the ordinary ones, making mirror matter invisible to us, and detectable just via their gravitational effects.
To this basic model of interactions, it is possible to add other interactions involving mixings between ordinary and mirror particles.
The most important of them is at present the kinetic mixing of photons, that in the mirror paradigm would be responsible of the positive results of the dark matter direct detection experiments.
Extensive reviews on mirror matter at astro- and particle physics levels can be found, for example, in Refs.\cite{Ciarcelluti:2010zz,Okun:2006eb,Ciarcelluti:2003wm,Foot:2003eq}.
The original idea and the first applications of mirror matter are present in Refs. \cite{Lee:1956qn,Blinnikov:1983gh,Khlopov:1989fj,Foot:1991bp}.

One of the key points in the macroscopic mirror theory is that, even if the physical laws are the same as ordinary matter, the initial conditions can be different.
This means that the densities of particles and their temperatures can be different, leading to the need of only two free parameters that describe the basic mirror model, defined as:
\begin{eqnarray} \label{mir-param}
  x &\equiv& \left( s' \over s \right)^{1/3} \approx \left( T' \over T \right) \;, \nonumber \\
  \beta &\equiv& \frac{\Omega'_{\rm b}}{\Omega_{\rm b}} \;,
\end{eqnarray}
where $s$ ($s'$), $T$ ($T'$) and $\Omega_{b}$ ($\Omega'_{b}$) are respectively the ordinary (mirror) entropy density, photon temperature, and cosmological baryon density.

Different cosmological parameters for initial temperatures and densities mean different cosmological evolutions for the two kinds of particles, in particular concerning the key phenomena of Big Bang nucleosynthesis (BBN), recombination, cosmic microwave background (CMB), large scale structure (LSS) formation, and the following evolution at lower scales, as galactic and stellar formation and evolution.
The CMB and LSS were well studied\cite{Ciarcelluti:2004ip,Ciarcelluti:2004ik,Ciarcelluti:2003wm,Berezhiani:2003wj,Ciarcelluti:2004ij,Ignatiev:2003js}, and a recent work \cite{Ciarcelluti:2012zz} has even shown that mirror matter can fit the observations with the same level of accuracy as generic cold dark matter (CDM).
Primordial nucleosynthesis was studied in the past in several works \cite{Berezhiani:2000gw,Ciarcelluti:2008vm,Ciarcelluti:2009da,Ciarcelluti:2010zz,Blinnikov:1983gh,Khlopov:1989fj}, obtaining historically the first bound on mirror matter parameters \cite{Berezhiani:1995am}.
In fact, if for example the temperature of the mirror particles would be the same as that of ordinary ones, the contribution of mirror relativistic species to the Hubble expansion rate at BBN epoch would be equivalent to that of an effective number of extra-(massless)neutrino families $\Delta N_\nu \simeq 6.14$.
This would be in conflict with any estimate, even the most conservative one.
Then, considering the weak bound $\Delta N_\nu \lesssim 1$ and just applying the approximate relation $\Delta N_\nu \approx 6.14 x^4$, one obtains $x \lesssim 0.7$ \cite{Berezhiani:1995am,Berezhiani:2000gw}.
The 4-th power of $x$ gives a mild dependence on this parameter.
In view of the definition \ref{mir-param} of $x$, this simply means that the mirror particles should have a lower temperature than the ordinary ones in the early Universe.

One of the peculiarities of mirror matter is that it not only influences the ordinary BBN, but it has his own mirror BBN!
This is a parallel primordial nucleosynthesis that is influenced by the ordinary baryonic matter, in an analogous way and for the same reason as mirror matter influences ordinary BBN.
The big difference is that, while ordinary BBN receives very low influence by the mirror matter, since $\Delta N_\nu \propto x^4$ and $x < 1$, the opposite happens for the mirror BBN, since instead for it $\Delta N_\nu \propto x^{-4}$!

As the mirror particles have a lower temperature than the ordinary ones, the conditions required to start the nucleosynthesis are reached at earlier cosmological times, which mean different conditions, and in particular a larger cosmic expansion rate.
This effect was studied in previous works, obtaining as a main result an increased production of mirror helium He$'$ in comparison with the ordinary one.
This last effect was studied also in presence of the photon-mirror photon kinetic mixing, obtaining similar results \cite{Ciarcelluti:2008qk,Ciarcelluti:2010dm}.
The primordial abundance $Y'$ of He$'$ is dependent on the inverse of the parameter $x$, and can reach very high values, up to $Y'= 0.8-0.9$ for low values $x \sim 0.1$.
All these studies considered valid the approximate relation $x \simeq T'/T$ expressed in Eq.\ref{mir-param}.
Indeed this relation is valid along most of the history of the Universe, but not at the period of BBN, since at this time the electron-positron annihilations in both ordinary and mirror sectors heat the respective photons at different times, inducing large differences between $x$ and $T'/T$, up to 30-40\%.
This effect and the thermodynamics of the early Universe were studied in details in Ref.\cite{Ciarcelluti:2008vs}.
Here we use the results of that work for the degrees of freedom and temperatures in the two particle sectors, in order to obtain a more detailed numerical description of the ordinary and mirror primordial nucleosynthesis processes, and more carefully predict the primordial abundances.
Some preliminary results of this analysis were previously presented in Refs.\cite{Ciarcelluti:2008vm,Ciarcelluti:2009da,Ciarcelluti:2010zz}.

An accurate study of BBN in ordinary and mirror particle sectors is important for the following reasons.
\begin{itemize}
\item At the present status of knowledge, the standard theory of BBN is essentially dependent on just one parameter, the baryonic asymmetry $\eta={n_{\rm b}}/{n_\gamma}$.
If compared with other cosmological tools, as for example CMB, that have much more free parameters, one can easily understand its importance as a key test for any dark matter candidate. 
\item In standard BBN there is the still open "lithium problem", related to the discrepancy between observations and predictions of the primordial abundance of this nuclide.
This suggests the need for new physics beyond the standard model, then it is important to understand if mirror matter can alleviate this problem.
\item The interpretation of DAMA and other direct detection experiments in terms of mirror matter is dependent on the abundance of mirror helium He$'$ and heavier elements (so called ``metals''), then a correct estimate of their abundances is crucial in order to verify this hypothesis \cite{Foot:2008nw,Foot:2012rk}.
\item The mirror BBN furnishes the primordial chemical composition of dark matter, that sets the initial conditions for the formation and evolution of structures at cosmic, galactic and subgalactic scales.
The effect of the enhanced abundance of He$'$ on the evolution of mirror stars has already been studied, showing a large effect \cite{Berezhiani:2005vv}.
It is fundamental to estimate also the abundances of metals, that are responsible of many opacitive processes of matter, involved in fragmentation processes forming galaxies and stars.
\end{itemize}



\begin{table*}
\caption{Elements produced in the ordinary sector. The last row includes all elements with atomic mass larger than 7.}
\begin{ruledtabular}
\begin{tabular}{lcccccccc}
                                & {standard}        & $ x=0.1 $         & $ x=0.2 $         & $ x=0.3 $         & $ x=0.4 $         & $ x=0.5 $         & $ x=0.6 $         & $ x=0.7 $   \\ \colrule
$n{\rm /H} \: (10^{-16})$       & 1.161\hphantom{0} & 1.161\hphantom{0} & 1.160\hphantom{0} & 1.159\hphantom{0} & 1.510\hphantom{0} & 1.505\hphantom{0} & 1.527\hphantom{0} & 2.044\hphantom{0} \\
$p$\hphantom{000000}            & 0.7518            & 0.7518            & 0.7516            & 0.7511            & 0.7495            & 0.7463            & 0.7409            & 0.7326 \\
${\rm D/H} \: (10^{-5})$        & 2.554\hphantom{0} & 2.555\hphantom{0} & 2.558\hphantom{0} & 2.575\hphantom{0} & 2.618\hphantom{0} & 2.709\hphantom{0} & 2.874\hphantom{0} & 3.144\hphantom{0} \\
${\rm T/H} \: (10^{-8})$        & 8.064\hphantom{0} & 8.065\hphantom{0} & 8.076\hphantom{0} & 8.132\hphantom{0} & 8.280\hphantom{0} & 8.588\hphantom{0} & 9.146\hphantom{0} & 10.07\hphantom{000} \\
${\rm ^3He/H} \: (10^{-5})$     & 1.038\hphantom{0} & 1.038\hphantom{0} & 1.038\hphantom{0} & 1.041\hphantom{0} & 1.046\hphantom{0} & 1.058\hphantom{0} & 1.080\hphantom{0} & 1.113\hphantom{0} \\
${\rm ^4He}$\hphantom{0000000}  & 0.2483            & 0.2483            & 0.2485            & 0.2491            & 0.2506            & 0.2538            & 0.2592            & 0.2675 \\
${\rm ^6Li/H} \: (10^{-14})$    & 1.111\hphantom{0} & 1.111\hphantom{0} & 1.113\hphantom{0} & 1.124\hphantom{0} & 1.151\hphantom{0} & 1.210\hphantom{0} & 1.318\hphantom{0} & 1.499\hphantom{0} \\
${\rm ^7Li/H} \: (10^{-10})$    & 4.549\hphantom{0} & 4.548\hphantom{0} & 4.543\hphantom{0} & 4.523\hphantom{0} & 4.468\hphantom{0} & 4.356\hphantom{0} & 4.162\hphantom{0} & 3.871\hphantom{0} \\
${\rm ^7 Be/H} \: (10^{-10})$   & 4.266\hphantom{0} & 4.266\hphantom{0} & 4.260\hphantom{0} & 4.238\hphantom{0} & 4.177\hphantom{0} & 4.051\hphantom{0} & 3.832\hphantom{0} & 3.502\hphantom{0} \\
${\rm ^8Li + /H} \: (10^{-15})$ & 1.242\hphantom{0} & 1.242\hphantom{0} & 1.243\hphantom{0} & 1.251\hphantom{0} & 1.269\hphantom{0} & 1.306\hphantom{0} & 1.370\hphantom{0} & 1.464\hphantom{0} \\
\end{tabular}
\end{ruledtabular}
\label{tab-ord-BBN}
\end{table*}

\section{Models}

I consider models in which the dark matter is made of mirror matter, and there are no interactions between ordinary and mirror particles except for the gravitation that links the two sectors.
Each set of particles is assumed to be in thermodynamical equilibrium (as usual in primordial nucleosynthesis studies) independently of the other one.
This implies that there are no entropy exchanges between the two sectors, then the entropy densities are separately conserved, and the parameter $x$ is constant by definition.
Using the definition of the entropy density \cite{Kolb:1990vq}, the parameter $x$ is 
\begin{equation} \label{xmir}
  x \equiv \left( s' \over s \right)^{1/3}
    = \left[q'(T') \over q(T) \right]^{1/3}{T' \over T}
~~,
\end{equation}
where $q(T)$ and $q'(T')$ are the ordinary and mirror entropic degrees of freedom.
The ratio of entropic degrees of freedom in the two sectors determines the mirror temperature as a function of the ordinary one, once the parameter $x$ is fixed.
This ratio is not constant at the times of primordial nucleosyntheses, since the $e^+$-$e^-$ annihilations heat the photons of each sectors independently, and they happen at different times in the early Universe.
This and related aspects were studied in details in Ref.\cite{Ciarcelluti:2008vs}.
Here we use the same treatment, valid for the ranges of temperatures which we are interested in (below $\sim$ 10 MeV).
It is based on the numerical solution of the equations for the conservations of ordinary and mirror entropies, and the absence of entropy exchanges (constancy of $x$):
\begin{eqnarray}\label{eqs:1-2-3}
&&\frac{\frac{7}{8}q_{e}(T)+q_{\gamma}}{\frac{7}{8}q_{\nu}} \left(\frac{T}{T_{\nu}}\right)^3 = \frac{22}{21} ~~, \nonumber\\
&&\frac{\frac{7}{8}q_{e}(T')+q_{\gamma}}{\frac{7}{8}q_{\nu}} \left(\frac{T'}{T_{\nu}'}\right)^3 = \frac{22}{21} ~~, \nonumber\\
&&\frac{\left[ \frac{7}{8}q_e(T')+ q_{\gamma} \right] T'^3+ \frac{7}{8} q_{\nu} 
T_{\nu}'\,^3} {\left[  \frac{7}{8}q_e(T)+ q_{\gamma} \right] T^3 + 
\frac{7}{8} q_{\nu} T_{\nu}^3} = x^3 ~~,
\end{eqnarray}
where $q_i$ is the entropic degrees of freedom of species $i$, $T_{\nu}$ and $T_{\nu}'$ are the temperatures of the ordinary and mirror neutrinos.

Since ordinary and mirror particles have the same microphysics, I assume that the neutrino decoupling temperature is the same in each sector, which means, due to the initial difference of temperatures, that the decouplings of ordinary and mirror neutrinos take place at different times in the early Universe.
This simplification is justified by the fact that, despite being the mirror sector colder, the final value of the neutrino to photon temperature ratio is the same in both sectors.

Together with $x$, the second free parameter for the mirror BBN is the mirror baryon to photon ratio (or baryonic asymmetry) $\eta' = n'_b / n'_\gamma$, which can be expressed in terms of the ordinary baryon to photon ratio $\eta = n_b / n_\gamma$ and the mirror parameters $x$ and $\beta$:
\begin{equation} \label{}
\eta' = \beta x^{-3} \eta > \eta ~~,
\end{equation}
where the inequality is due to the bounds on $x$ and the expected values for $\beta > 1$.

Since the nuclear physics is the same for ordinary and mirror matter, it is possible to use and modify a pre-existing code for primordial nuceosynthesis, that numerically solves the equations governing the production and evolution of nuclides.
The choice is the well-tested and fast Wagoner-Kawano code \cite{Wagoner:1972jh,Kawano:1992ua}, which has enough accuracy for the purposes of this analysis.
The numerical code has been doubled to include the mirror sector, and modified in order to take into account the evolution of the temperature of the mirror particles and the degrees of freedom of both sectors, according to the aforementioned treatment \cite{Ciarcelluti:2008vs}.
For the neutron lifetime we consider the value $\tau = 885.7$ s, while for the final baryon to photon ratio $\eta = 6.14 \cdot 10^{-10}$.
We consider the usual standard number of neutrino families for ordinary matter $N_{\nu}=3.04$.
Then, the only two free parameters of the code are the mirror ones, $x$ and $\beta$.
Several models are computed for $x$ ranging from 0.1 to 0.7 and $\beta$ from 1 to 5, that are the values of cosmological interests.

\begin{table*}
\caption{Elements produced in the mirror sector. The last row includes all elements with atomic mass larger than 7.}
\begin{ruledtabular}
\begin{tabular}{llllllll}
  & $ x=0.1\, (\beta = 5) $
  & $ x=0.2\, (\beta = 5) $
  & $ x=0.3\, (\beta = 5) $ 
  & $ x=0.4\, (\beta = 5) $
  & $ x=0.5\, (\beta = 5) $
  & $ x=0.6\, (\beta = 5) $
  & $ x=0.7\, (\beta = 5) $ \\ \colrule
$n{\rm /H} $     &\ 5.762 $\cdot 10^{-25}$ &\ 2.953 $\cdot 10^{-24}$ &\ 2.590 $\cdot 10^{-22}$ &\ 2.908 $\cdot 10^{-21}$ &\ 1.840 $\cdot 10^{-20}$ &\ 6.858 $\cdot 10^{-20}$ &\ 1.726 $\cdot 10^{-19}$ \\
$p $             &\ 0.1735                 &\ 0.2840                 &\ 0.3646                 &\ 0.4357                 &\ 0.4966                 &\ 0.5488           &\ 0.5924                       \\
${\rm D/H} $     &\ 1.003 $\cdot 10^{-12}$ &\ 3.090 $\cdot 10^{-10}$ &\ 4.838 $\cdot 10^{-9}$  &\ 2.240 $\cdot 10^{-8}$  &\ 6.587 $\cdot 10^{-8}$  &\ 1.553 $\cdot 10^{-7}$  &\ 3.279 $\cdot 10^{-7}$  \\
${\rm T/H}$      &\ 9.679 $\cdot 10^{-21}$ &\ 4.999 $\cdot 10^{-16}$ &\ 1.238 $\cdot 10^{-13}$ &\ 2.603 $\cdot 10^{-12}$ &\ 2.108 $\cdot 10^{-11}$ &\ 1.030 $\cdot 10^{-10}$ &\ 3.722 $\cdot 10^{-10}$ \\
${\rm ^3He/H}$   &\ 3.282 $\cdot 10^{-6}$  &\ 3.522 $\cdot 10^{-06}$ &\ 3.740 $\cdot 10^{-6}$  &\ 3.949 $\cdot 10^{-6}$  &\ 4.172 $\cdot 10^{-6}$  &\ 4.415 $\cdot 10^{-6}$  &\ 4.691 $\cdot 10^{-6}$  \\
${\rm ^4He} $    &\ 0.8051                 &\ 0.7233                 &\ 0.6351                 &\ 0.5648                 &\ 0.5035                 &\ 0.4514           &\ 0.4077                       \\
${\rm ^6Li/H}$   &\ 7.478 $\cdot 10^{-21}$ &\ 1.241 $\cdot 10^{-18}$ &\ 1.309 $\cdot 10^{-17}$ &\ 4.460 $\cdot 10^{-17}$ &\ 1.016 $\cdot 10^{-16}$ &\ 1.923 $\cdot 10^{-16}$ &\ 3.361 $\cdot 10^{-16}$ \\
${\rm ^7Li/H} $  &\ 1.996 $\cdot 10^{-7}$  &\ 7.162 $\cdot 10^{-8}$  &\ 3.720 $\cdot 10^{-8}$  &\ 2.289 $\cdot 10^{-8}$  &\ 1.535 $\cdot 10^{-8}$  &\ 1.086 $\cdot 10^{-8}$  &\ 7.962 $\cdot 10^{-9}$  \\
${\rm ^7 Be/H}$  &\ 1.995 $\cdot 10^{-7}$  &\ 7.159 $\cdot 10^{-8}$  &\ 3.675 $\cdot 10^{-8}$  &\ 2.236 $\cdot 10^{-8}$  &\ 1.489 $\cdot 10^{-8}$  &\ 1.041 $\cdot 10^{-8}$  &\ 7.885 $\cdot 10^{-9}$  \\
${\rm ^8Li +/H}$ &\ 4.354 $\cdot 10^{-9}$  &\ 3.458 $\cdot 10^{-10}$ &\ 5.926 $\cdot 10^{-11}$ &\ 1.396 $\cdot 10^{-11}$ &\ 3.827 $\cdot 10^{-12}$ &\ 1.168 $\cdot 10^{-12}$ &\ 3.949 $\cdot 10^{-13}$ \\ \botrule
  & $ x=0.1\, (\beta = 1) $
  & $ x=0.2\, (\beta = 1) $
  & $ x=0.3\, (\beta = 1) $ 
  & $ x=0.4\, (\beta = 1) $
  & $ x=0.5\, (\beta = 1) $
  & $ x=0.6\, (\beta = 1) $
  & $ x=0.7\, (\beta = 1) $ \\ \colrule
$n/H $           &\ 8.888 $\cdot 10^{-17}$ &\ 1.110 $\cdot 10^{-16}$ &\ 1.915 $\cdot 10^{-16}$ &\ 1.620 $\cdot 10^{-16}$ &\ 2.058 $\cdot 10^{-16}$ &\ 1.399 $\cdot 10^{-16}$ &\ 2.076 $\cdot 10^{-16}$ \\
$p $             &\ 0.1772                 &\ 0.2831                 &\ 0.3675                 &\ 0.4400                 &\ 0.5028                 &\ 0.5566                 &\ 0.6017                       \\
$D/H $           &\ 1.331 $\cdot 10^{-6}$  &\ 4.086 $\cdot 10^{-6}$  &\ 7.094 $\cdot 10^{-6}$  &\ 1.018 $\cdot 10^{-5}$  &\ 1.352 $\cdot 10^{-5}$  &\ 1.743 $\cdot 10^{-5}$  &\ 2.235 $\cdot 10^{-5}$  \\
$T/H$            &\ 3.068 $\cdot 10^{-9}$  &\ 1.192 $\cdot 10^{-8}$  &\ 2.190 $\cdot 10^{-8}$  &\ 3.228 $\cdot 10^{-8}$  &\ 4.358 $\cdot 10^{-8}$  &\ 5.675 $\cdot 10^{-8}$  &\ 7.328 $\cdot 10^{-8}$  \\
${\rm ^3He/H}$   &\ 5.228 $\cdot 10^{-6}$  &\ 6.119 $\cdot 10^{-6}$  &\ 6.880 $\cdot 10^{-6}$  &\ 7.566 $\cdot 10^{-6}$  &\ 8.232 $\cdot 10^{-6}$  &\ 8.931 $\cdot 10^{-6}$  &\ 9.719 $\cdot 10^{-6}$  \\
${\rm ^4He}$     &\ 0.8226                 &\ 0.7168                 &\ 0.6326                 &\ 0.5602                 &\ 0.4974                 &\ 0.4436                 &\ 0.3984                 \\
${\rm ^6Li/H}$   &\ 8.638 $\cdot 10^{-15}$ &\ 1.422 $\cdot 10^{-14}$ &\ 1.660 $\cdot 10^{-14}$ &\ 1.747 $\cdot 10^{-14}$ &\ 1.790 $\cdot 10^{-14}$ &\ 1.845 $\cdot 10^{-14}$ &\ 1.951 $\cdot 10^{-14}$ \\
${\rm ^7Li/H}$   &\ 5.712 $\cdot 10^{-8}$  &\ 1.867 $\cdot 10^{-8}$  &\ 8.930 $\cdot 10^{-9}$  &\ 4.953 $\cdot 10^{-9}$  &\ 2.948 $\cdot 10^{-9}$  &\ 1.811 $\cdot 10^{-9}$  &\ 1.120 $\cdot 10^{-9}$  \\
${\rm ^7 Be/H}$  &\ 5.711 $\cdot 10^{-8}$  &\ 1.863 $\cdot 10^{-8}$  &\ 8.878 $\cdot 10^{-9}$  &\ 4.896 $\cdot 10^{-9}$  &\ 2.891 $\cdot 10^{-9}$  &\ 1.755 $\cdot 10^{-9}$  &\ 1.064 $\cdot 10^{-9}$  \\
${\rm ^8Li +/H}$ &\ 2.036 $\cdot 10^{-10}$ &\ 1.468 $\cdot 10^{-11}$ &\ 2.514 $\cdot 10^{-12}$ &\ 5.944 $\cdot 10^{-13}$ &\ 1.657 $\cdot 10^{-13}$ &\ 5.184 $\cdot 10^{-14}$ &\ 1.814 $\cdot 10^{-14}$ \\
\end{tabular}
\end{ruledtabular}
\label{tab-mir-BBN}
\end{table*}

\section{Results}

For each couple of mirror parameters ($x$, $\beta$) the primordial abundances of both ordinary and mirror elements are derived.
The ordinary BBN, as expected, is independent on the density of mirror baryons (mirror baryonic asymmetry), then it depends, once fixed the microphysical parameters and the ordinary baryonic asymmetry, on just one parameter, the ratio of entropies $x$.
The mirror BBN, instead, is clearly dependent on both $x$ and the cosmic mirror baryonic density, expressed by the parameter $\beta$.

\begin{figure}
\includegraphics[scale=0.5]{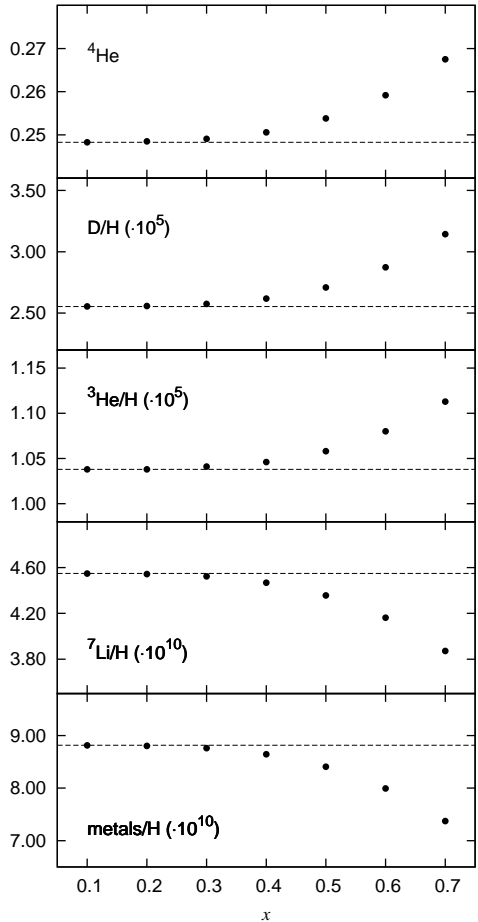}
\caption{{\it Primordial abundances of ordinary $^4He$, $D$, $^3He$, $^7Li$ and metals (elements heavier than $^4He$) for several values of $x$ and compared with the predictions of the standard model (dashed lines).
}}
\label{fig-ord-BBN}
\end{figure}

In Table \ref{tab-ord-BBN} the primordial abundances of elements produced by ordinary nucleosynthesis, for different values of $x$ compared with the standard model of nucleosynthesis (absence of dark matter), are reported.
Protons and ${\rm ^4He}$ are expressed in mass fraction, all the others in ratios to the proton abundance.
In the last row, indicated with ${\rm ^8Li +}$, the contributions of the elements with atomic mass larger than 7 are included all together.
The evolution of the abundances has been followed until the end of BBN process (at $T \sim 8\cdot 10^{-4}$ MeV).
It is evident that the differences with the standard BBN appear only for $x>0.1$ (for $x=0.1$ they are of order $10^{-4}$ or less), and for $x=0.3$ they are limited to below 1\%, but they increase for increasing values of $x$.
The abundances of most elements (D, T, ${\rm ^3He}$, ${\rm ^4He}$, ${\rm ^6Li}$, ${\rm ^8Li +}$) increase with $x$, while those of ${\rm ^7Li}$ and ${\rm ^7 Be}$ decrease.
This predicted decrease for ${\rm ^7Li}$ is an interesting result, since it goes exactly in the direction required to solve the still open ``lithium problem''.
At first sight, the entity of the decrease should not be sufficient to solve this problem of standard BBN, but could certainly alleviate it.
A dedicated statistical analysis will help to better evaluate this interesting possibility carried by the mirror matter.
The trends of the observable primordial abundances, namely ${\rm ^4He}$, D, ${\rm ^3He}$, ${\rm ^7Li}$ and metals (the sum of the abundances of all elements heavier than ${\rm ^4He}$) are plotted in Figure \ref{fig-ord-BBN} as functions of $x$ and compared with the standard model.
As expected, the trend with $x$ is not linear, since it is (indirectly) dependent on the ordinary degrees of freedom, that scale as $x^4$.
This dependence is the reason of the negligible effects predicted at lower $x$.

\begin{figure}
\includegraphics[scale=0.5]{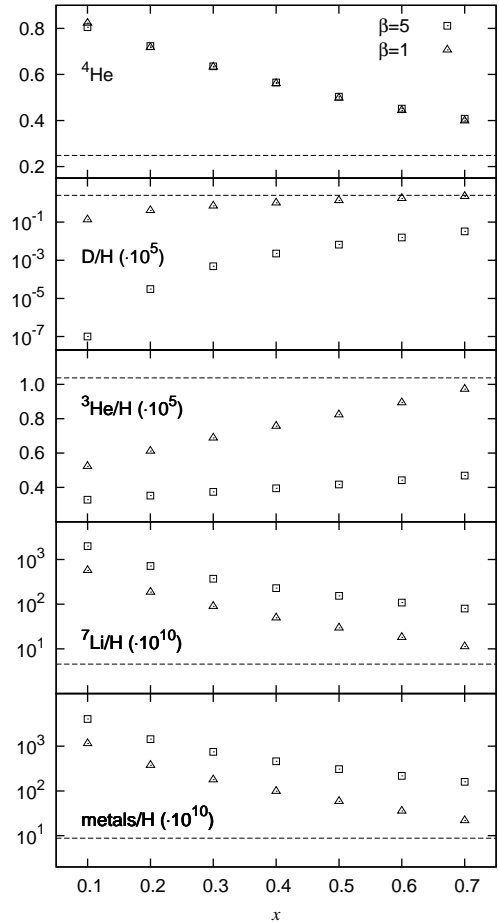}
\caption{{\it Primordial abundances of mirror $^4He$, $D$, $^3He$, $^7Li$ and metals (elements heavier than $^4He$) for several values of $x$ and two different $\beta$, and compared with the predictions of the standard model (dashed lines).
}}
\label{fig-mir-BBN}
\end{figure}

\begin{figure}
\includegraphics[scale=0.44]{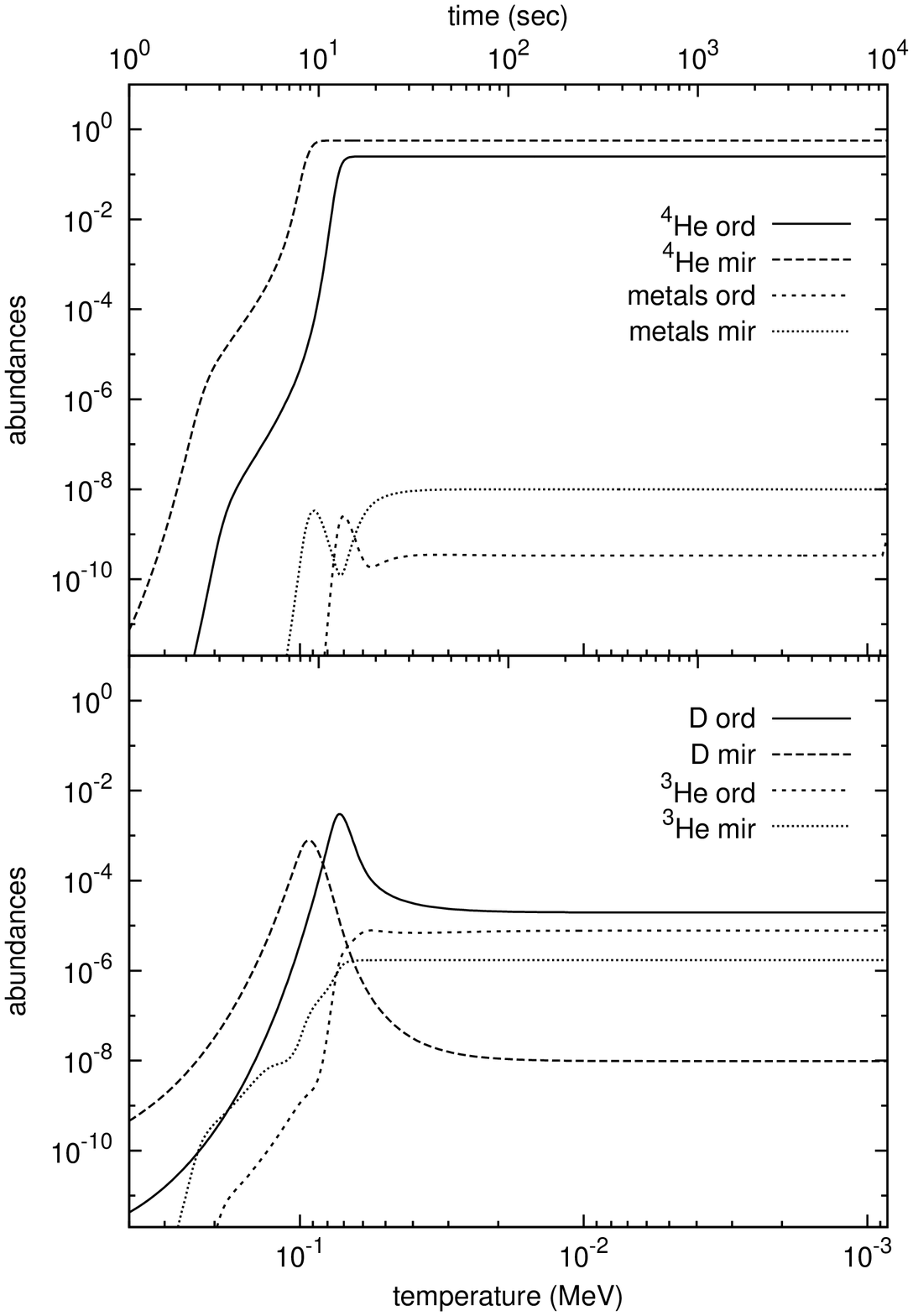}
\caption{{\it Time/temperature evolution of nuclides during ordinary and mirror primordial nucleosyntheses. 
The models have the mirror parameters $x=0.4$ and $\beta = 5$.}}
\label{std_ord_nucl}
\end{figure}

The results of the models for mirror nuclei are shown in Table \ref{tab-mir-BBN}, that is the analogous of Table \ref{tab-ord-BBN}.
Since in this case the models depend on both the ratio of entropies and the ratio of baryonic densities, they are computed for the same different values of $x$ as for the ordinary BBN, and for two different values of $\beta$, chosen at the extremes of the range (1 and 5) in order to maximize the effects of their change.
One immediately sees that the mirror BBN is very different from the ordinary one.
This is what in fact one expects, since the contribution of the ordinary particles to the mirror degrees of freedom has a dependence as $x^{-4}$, then it is significant and becomes higher for lower $x$.
As expected, for higher values of $x$ the primordial abundances of mirror nuclides become less different from the ordinary ones, since the temperature of the mirror particles becomes higher, and then similar to that of the ordinary ones, in view of the approximate relation $T' \sim xT$.
In addition, the same general trend is observed for lower values of $\beta$, that means baryonic densities similar to the ordinary ones.
It is not simple to describe the trends of the mass fractions by changing sectors and parameters, as the final abundances depend on many physical processes acting together, but one can try to summarize some results.
Comparing the mirror nuclei with the ordinary ones, one observes: much less residual neutrons and considerably less protons, that essentially went to build ${\rm ^4He}$ nuclei; 
much more ${\rm ^4He}$ (clearly the dependence on $x$ is the opposite as for the protons); 
several orders of magnitude less D, T, ${\rm ^3He}$; 
much less ${\rm ^6Li}$ for $\beta=5$ and similar abundances for $\beta=1$;
much more ${\rm ^7Li}$, ${\rm ^7 Be}$ and ${\rm ^8Li +}$.
Considering the trends with $x$, mirror abundances of $n$, $p$, ${\rm ^4He}$ and ${\rm ^8Li +}$ have opposite trends than the corresponding ordinary elements.
Comparing the predictions obtained for the different values of $\beta$, one observes the following:
the trends with $x$ are the same for each element; 
the abundances of ${\rm ^4He}$ are very similar;
the abundances of ${\rm ^3He}$ become almost double going from $\beta=5$ to $\beta=1$;
for the lower $\beta$ there is much more D, T and ${\rm ^6Li}$ (some orders of magnitude) and much less ${\rm ^7Li}$, ${\rm ^7 Be}$ and ${\rm ^8Li +}$ (around one order).
Analogously to what done for the ordinary matter, in Figure \ref{fig-mir-BBN} I plot the abundances of mirror ${\rm ^4He}$, D, ${\rm ^3He}$, ${\rm ^7Li}$ and metals, as functions of $x$ and for the two values of $\beta$.
The previously mentioned trends, namely the growing similarity with the standard model abundances for higher $x$ and lower $\beta$, are evident.
Differently from the ordinary nuclei, the mirror ones are not directly observable, but their primordial abundances are a key ingredient for studies of the following evolution of the Universe at all scales, and for the aforementioned interpretation of the dark matter direct detection experiments.
The computed mass fraction for mirror ${\rm ^4He}$ is in qualitative agreement with what predicted by previous analytical studies \cite{Berezhiani:2000gw,Ciarcelluti:2010zz}, confirming that it is larger that the ordinary one for every $x$ and becomes the dominant mass contribution for $x \lesssim 0.5$, meaning that dark matter would be dominated by mirror helium.
Another important result of the simulations is the prediction of a much larger abundance of metals produced by mirror nucleosynthesis.
These elements have a large influence on the opacity of mirror matter, that has an important role in many astrophysical processes, as for example the fragmentation of primordial gas during the phase of contraction.

In order to complete the analysis, I show in Figure \ref{std_ord_nucl} the evolution of the abundances of ordinary and mirror D, ${\rm ^3He}$, ${\rm ^4He}$ and metals.
The models used have the parameters $x=0.4$ and $\beta=5$.
The evolutions of standard and ordinary abundances are very similar (and for this reason the standard ones are not shown in figure), while the mirror ones have a similar shape, but different values.
In particular, they appear shifted towards earlier times, as a consequence of the smaller temperature of the primordial mirror plasma.


\section{Conclusions}

In this paper I present the detailed study of the primordial nucleosynthesis in presence of mirror dark matter, in its basic model with only gravitational interactions between ordinary and mirror particles.
The BBN is studied for both kinds of matter, using an accurate treatment of the thermodynamics of the early Universe, based on the work done in Ref. \cite{Ciarcelluti:2008vs}, which considers the changes in the radiation temperatures due to the two $e^+$-$e^-$ annihilations.
The present analysis shows the results of accurate numerical simulations and updates all the previous works.
Fixing the cosmological parameters to their standard values, only the two free mirror parameters are considered, namely the ratio of entropies $x$ and the ratio of baryonic densities $\beta$.
Both ordinary and mirror nucleosynthesis are followed until their ends, obtaining the evolution and final abundances of primordial elements in every sector.
For the ordinary nuclides, they depend only on the parameter $x$, while for the mirror ones they are dependent also on $\beta$.
As expected, the upper bound $x<0.7$ limits the effect of mirror particles on ordinary nucleosynthesis, that is negligible for $x=0.1$ and starts to be around few percent for $x=0.3$, with a dependence growing with $x$.
An interesting unexpected result is the prediction of a lower abundance of ${\rm ^7Li}$.
This effect could help to alleviate the ``lithium problem'', but it requires a future dedicated statistical analysis.
In the mirror sector, the Big Bang nucleosynthesis produces in a similar way mirror nuclides, but with different abundances as a consequence of its different initial conditions.
In particular, as previously analytically predicted, there is an enhanced production of mirror ${\rm ^4He}$, that becomes the dominant nuclide for $x \lesssim 0.5$, and arrives at more than $80\%$ for the lowest values of $x$.
This effect has a very small dependence on $\beta$.
In addition, there is a much larger (few orders of magnitude) production of mirror metals (elements heavier that ${\rm ^4He}$).
Even if their abundances are anyway very low, they could have consequences on the opacity of dark matter, and on its many related astrophysical phenomena.
This work provides the primordial chemical composition of the mirror dark matter, that has to be used in studied of the evolution of the Universe at all scales, and in the analyses of the dark matter direct detection experiments.




\bibliography{mir_bbn}

\end{document}